\documentclass[conference]{IEEEtran}
\IEEEoverridecommandlockouts
\usepackage{cite}
\usepackage{amsmath,amssymb,amsfonts}
\usepackage{algorithmic}
\usepackage{graphicx}
\usepackage{textcomp}
\usepackage{xcolor}
\usepackage{float}
\usepackage{stfloats}
\usepackage{array}
\usepackage[hidelinks]{hyperref}
\usepackage{multirow}
\def\BibTeX{{\rm B\kern-.05em{\sc i\kern-.025em b}\kern-.08em
    T\kern-.1667em\lower.7ex\hbox{E}\kern-.125emX}}
\begin{document}

\title{AMECxSV: Adaptive Metadata-Driven Embedding-Fusion Calibration for X-Lingual Speaker Verification}

\author{
\IEEEauthorblockN{Xin Wei$^{1,*}$, Shi He$^{1,*}$, Yihe Yuan$^1$, Huang-Cheng Chou$^2$,\\
Sudarsana Reddy Kadiri$^2$, and Shrikanth Narayanan$^2$}
\IEEEauthorblockA{
$^1$University of Southern California, USA\\
$^2$Signal Analysis and Interpretation Laboratory (SAIL), University of Southern California, USA\\
\{xwei6575, shihe, yiheyuan, chouhuan, skadiri\}@usc.edu, shri@sipi.usc.edu
}
}

\maketitle
\begingroup
\renewcommand{\thefootnote}{*}
\footnotetext{These authors contributed equally.}
\endgroup

\begin{abstract}
In X-lingual automatic speaker verification (ASV), fixed front-end scores vary in reliability with language match, duration, and score source. We propose AMECxSV, an adaptive metadata-driven embedding-fusion calibration backend for metadata-available settings. AMECxSV fuses trial scores with metadata to produce calibrated target posteriors, with optional posterior-confidence abstention; metadata serve as calibration context, not speaker evidence. 
On a development-derived speaker-disjoint held-out split, score+metadata heads reduce equal error rate (EER) from $3.15\%$ to $2.42\%$ for the official TidyVoice score source and from $0.64\%$ to $0.43\%$ for LI-MSV; the dual-score head reaches $0.43\%$ full-coverage EER. 
At $0.79$ coverage, abstention yields $0.03\%$ accepted-trial EER, not a full-coverage metric. 
Matched score-only, metadata-permutation, and metadata-only controls support calibration-context interpretation and limit claims to metadata-available scoring.
\end{abstract}


\section{Introduction}
Modern ASV front ends such as the x-vector, ECAPA-TDNN, ResNet, CAM++, ERes2NetV2 and WavLM have improved speaker-discriminative representations~\cite{snyder2018xvectors,desplanques2020ecapa,wang2023wespeaker,wang2023campp,chen2024eres2netv2,chen2021wavlm}. 
However, a stronger embedding extractor does not make every backend score equally reliable. Score reliability still varies with language, utterance duration, recording condition, and front-end design. 
We therefore study adaptive embedding-fusion calibration while keeping the embedding extractors fixed.

This issue is central in X-lingual ASV, including multilingual and cross-lingual trial conditions. 
Language mismatch can shift score distributions and short or unbalanced utterances can reduce score reliability~\cite{thienpondt2022crosslingual,mandasari2013qmf,mandasari2015quality}. Recent TidyVoice and TidyVoiceX-ASV protocols further emphasize speaker verification under multilingual trial conditions~\cite{farhadipour2026tidyvoice,farhadipour2026tidyvoicechallenge}. 
Their official evaluation phase withholds utterance-language information to assess language-independent systems, whereas the development protocol exposes language conditions for analysis. 
We therefore study a controlled metadata-available calibration setting, not a language-blind final-evaluation submission. Language and duration cues should not be treated as standalone speaker-verification evidence, because the task remains same-speaker versus different-speaker verification. 
Instead, they can provide calibration context for interpreting score reliability under a given trial condition.

To address the above-mentioned challenges, we propose AMECxSV, which is a metadata-driven embedding-fusion calibration framework for metadata-available X-lingual ASV. 
It operates on scores from fixed embedding systems rather than updating the embedding extractors. 
In the experiments, AMEC-FC and AMEC-ABS denote the full-coverage and confidence-reject variants of AMECxSV.

In summary, AMECxSV makes four contributions: (i) formalizing metadata-available X-lingual ASV reliability as embedding-fusion calibration, (ii) proposing AMEC-FC as a lightweight MLP backend for full-coverage score calibration, (iii) extending it to AMEC-ABS through posterior-confidence rejection, and (iv) validating AMECxSV with internal controls, ablations, coverage-risk analysis, external fixed-front-end experiments, and sanity checks. On the development-derived held-out split, AMECxSV score+metadata heads reduce full-coverage EER from 3.15\% to 2.42\% for the official TidyVoice score source and from 0.64\% to 0.43\% for LI-MSV. The dual-score head reaches 0.43\% full-coverage EER. Its confidence-reject variant reaches 0.03\% accepted-trial EER at 0.79 coverage, reported only as a selective-decision result. Internal controls show that metadata help only when aligned with score evidence.
The project website is available\footnote{\url{https://anonymous.4open.science/w/AMECxSV-6DB2/}}, including codebases and database.

\section{Related Work}
Classical ASV score calibration represents verification outputs as log-likelihood ratios and evaluates them with calibration and detection metrics such as $C_{\mathrm{llr}}$, detection cost functions (DCF), and detection error trade-off (DET) curves~\cite{martin1997det,brummer2006application}. 
Logistic calibration, PAV/isotonic calibration, and the BOSARIS toolkit provide standard foundations for calibration, score fusion, and cost-sensitive ASV evaluation~\cite{brummer2013bosaris,brummer2013pav,brummer2014comparison}. 
These methods make scores more interpretable and improve operating-point behavior, but a single global calibration cannot adjust to each trial's language or duration conditions.

Additionally, quality-aware and condition-aware calibration add trial side information to this framework. 
Quality measure functions (QMFs) model duration and noise effects in calibration, and later studies examined robustness under short-duration or low-SNR conditions~\cite{mandasari2013qmf,mandasari2015quality,nautsch2016robustness}. 
Trial-based calibration, robust condition-aware backends, and generative condition-aware calibration further adapt scores under heterogeneous or unseen conditions~\cite{lei2014trialcalibration,ferrer2021robustbackend,borgstrom2023generative}. 
Much of this work still focuses on one score stream or on condition models that are not tied to multi-front-end score evidence.

Multi-system and multi-score fusion are also common in ASV. Ordinary fusion combines score streams through logistic regression or BOSARIS-style calibration/fusion, and has been effective in challenge systems with quality-aware calibration~\cite{brummer2013bosaris,thienpondt2021idlabvoxsrc20}. 
However, ordinary score fusion does not explicitly specify how language-match and duration-reliability cues should shape the calibrated decision. 
AMECxSV is positioned as metadata-driven embedding-fusion calibration: score evidence from fixed embedding systems remains central, while metadata condition how that evidence is calibrated.

Reliable deployment also requires deciding when an automatic verification decision should be deferred. 
Reject-option and selective-classification methods trade coverage for lower accepted-set risk~\cite{geifman2017selective}, and fail-safe speaker recognition has studied trial-based calibration with rejection under calibration mismatch~\cite{ferrer2022failsafe}. 
AMECxSV keeps the ASV decision binary. 
Abstention is a confidence-control layer on top of the calibrated posterior, not a third ASV class.

\begin{figure*}[!t]
    \centering
    \includegraphics[width=\textwidth]{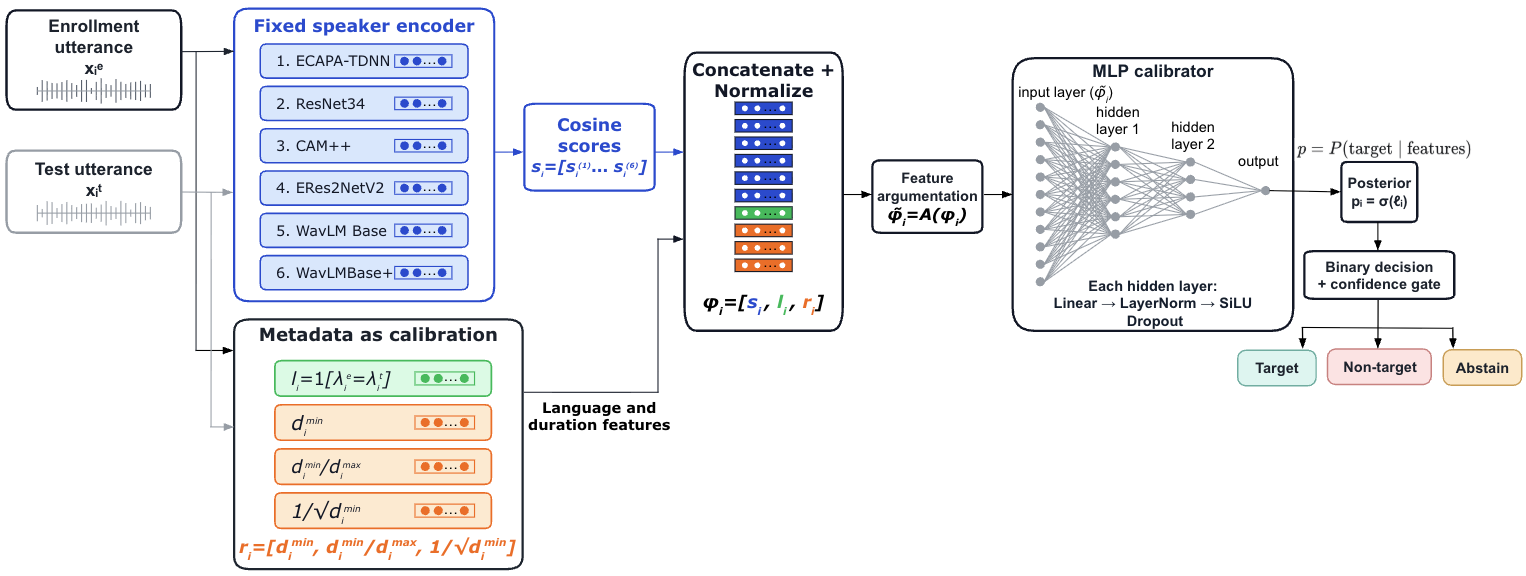}
    \caption{Overview of AMECxSV pipeline: adaptive metadata-driven embedding-fusion calibration backend for fixed embedding front ends.}
    \label{fig:amec_overview}
    \vspace{-3mm}
\end{figure*}

\section{Methodology}
\subsection{Task Formulation (Fixed-Extractor Problem)}
For trial $i$, define
\begin{equation}
    T_i=(x_i^e,x_i^t,t_i),
\end{equation}
where $x_i^e$ is the enrollment utterance, $x_i^t$ is the test utterance, and $t_i\in\{0,1\}$ is the binary ASV label.

Let $f^{(k)}$ denote the $k$th fixed speaker front end. It maps the enrollment and test utterances to embeddings
\begin{equation}
    \mathbf{z}_{i,e}^{(k)}=f^{(k)}(x_i^e),\qquad
    \mathbf{z}_{i,t}^{(k)}=f^{(k)}(x_i^t),
\end{equation}
from which a trial score $s_i^{(k)}$ is computed using cosine scoring or the fixed score backend provided with the front end. All $f^{(k)}$ are fixed; AMECxSV learns only the embedding-fusion calibration and decision function.

Figure~\ref{fig:amec_overview} summarizes the AMECxSV pipeline. Fixed speaker encoders produce trial-level scores, metadata enter as calibration context, and deterministic feature expansion feeds an MLP calibration backend. The backend output is converted to a calibrated posterior and then to a binary decision, with an optional confidence gate.

\subsection{Trial-Level Score and Metadata Representation}
The score vector ($\mathbf{s}$) for trial $i$ is
\begin{equation}
    \mathbf{s}_i=[s_i^{(1)},\ldots,s_i^{(K)}].
\end{equation}
Here, $K=6$ for the internal six-front-end experiment, $K=1$ for single external baseline experiments, and $K=2$ for dual-score external AMEC.

Language-match metadata are represented as
\begin{equation}
    L_i=\mathbb{I}[\lambda_i^e=\lambda_i^t],
\end{equation}
where $\lambda_i^e$ and $\lambda_i^t$ are the enrollment and test languages. Trials with $L_i=1$ are same-language trials, and trials with $L_i=0$ are cross-language trials. Directed language-pair information is retained for protocol statistics and analysis, while the compact AMEC input uses $L_i$.

Duration-reliability metadata are defined by
\begin{equation}
    d_i^{\min}=\min(d_i^e,d_i^t),\qquad
    d_i^{\max}=\max(d_i^e,d_i^t),
\end{equation}
\begin{equation}
    \mathbf{r}_i=
    \left[
    d_i^{\min},
    \frac{d_i^{\min}}{d_i^{\max}},
    \frac{1}{\sqrt{d_i^{\min}}}
    \right].
\end{equation}
The minimum duration captures the weakest utterance in the trial, the duration ratio captures enrollment-test imbalance, and the inverse-square-root term emphasizes short-duration risk.

The metadata vector is $\mathbf{m}_i=[L_i,\mathbf{r}_i]$, and the initial trial representation is
\begin{equation}
    \boldsymbol{\phi}_i=[\mathbf{s}_i,\mathbf{m}_i].
\end{equation}
Metadata condition calibration but do not replace score evidence (language-unavailable variants are noted in Appendix~\ref{sec:appendix_external_deployment}).

\subsection{Metadata-Driven Embedding-Fusion Calibration Model}
The AMECxSV calibration head applies a deterministic feature expansion
\begin{equation}
    \mathbf{x}_i=\mathcal{R}(\boldsymbol{\phi}_i).
\end{equation}
The expansion includes score summaries, order statistics, pairwise score differences and products, score-metadata interactions, and squared metadata terms. Implementation-specific feature dimensions are reported with the experiment details.

The expanded features are normalized as
\begin{equation}
    \tilde{\mathbf{x}}_i=\operatorname{Norm}(\mathbf{x}_i),
    \qquad
    u_i=h_\theta(\tilde{\mathbf{x}}_i),
\end{equation}
where the normalization statistics are computed on the calibration training split and $h_\theta$ is the MLP backend.

The calibrated logit and calibrated target posterior are
\begin{equation}
    \ell_i=\alpha u_i+\beta,\qquad p_i=\sigma(\ell_i),
\end{equation}
where $\alpha$ and $\beta$ are fitted on the validation split.
This affine step calibrates posterior scale. It is not the source of ranking gains; changes in EER arise from the learned score fusion and score-metadata interaction features before affine calibration.

The MLP is trained with balanced binary cross-entropy,
\begin{equation}
    \mathcal{L}_{\mathrm{cal}}
    =
    -\sum_{i=1}^{N} w_i
    \left[
    t_i \log \sigma(u_i)
    +
    (1-t_i)\log(1-\sigma(u_i))
    \right],
\end{equation}
where $w_i$ balances target and non-target trials. No abstention class is trained.

\subsection{Binary Decision and Confidence-based Abstention}
The full-coverage binary decision is
\begin{equation}
    y_i =
    \begin{cases}
    \mathrm{target}, & \ell_i \geq \tau,\\
    \mathrm{non\mbox{-}target}, & \ell_i < \tau.
    \end{cases}
\end{equation}
For actDCF, $\tau=\log((1-P_{\mathrm{tar}})/P_{\mathrm{tar}})$ is the fixed application threshold; validation tuning is used only for the abstention coverage gate.
AMEC-FC returns a target/non-target decision for every trial.

Posterior confidence is
\begin{equation}
    \kappa_i=\max(p_i,1-p_i).
\end{equation}
When $p_i$ is calibrated, $\kappa_i$ is the posterior confidence assigned to the chosen binary label.
For abstention, the acceptance mask is
\begin{equation}
    m_i=\mathbb{I}[\kappa_i\geq\eta_\kappa],
    \qquad
    \eta_\kappa=\operatorname{Quantile}_{1-\Gamma}\{\kappa_j:j\in\mathcal{V}\},
\end{equation}
where $\eta_\kappa$ is chosen on the validation set $\mathcal{V}$ for target coverage $\Gamma$.

The realized coverage is
\begin{equation}
    \mathrm{Coverage}=\frac{1}{N}\sum_{i=1}^{N}m_i.
\end{equation}
Accepted-set metrics must always be interpreted together with coverage. AMEC-ABS is AMEC-FC with this posterior-confidence gate.

\noindent\textbf{Variant summary.}
AMEC-FC returns a binary target/non-target decision for every trial. AMEC-ABS uses the same calibrated backend but accepts only trials that pass the validation-tuned confidence gate.

\section{Experimental Setup}
\subsection{Datasets and Speaker-Disjoint Splits}
The main experiments use the TidyVoiceX-ASV trial-level development protocol~\cite{farhadipour2026tidyvoice,farhadipour2026tidyvoicechallenge}. 
Each trial has one enrollment utterance and one test utterance. 
Target trials are same-speaker trials, and non-target trials are different-speaker trials. 
The speaker label is separate from the language condition.

For each trial, we retain enrollment language $\lambda_i^e$, test language $\lambda_i^t$, directed language pair $(\lambda_i^e,\lambda_i^t)$, language-match $L_i$, enrollment duration $d_i^e$, and test duration $d_i^t$. 
These fields align with the notation in Section~II.
This differs from the official TidyVoice evaluation phase, which withholds utterance-language information. We test metadata-available calibration, not language-blind challenge scoring.

The full TidyVoiceX-ASV development protocol contains 12M trials from 808 speakers, including 4M target and 8M non-target trials. 
It covers 39 languages and 1520 directed language pairs, with 6M same-language and 6M cross-language trials.
We split the development protocol into deterministic speaker-disjoint calibration, validation, and held-out test partitions. This held-out partition is not the official evaluation set. Target trials remain in their speaker partition. Non-target trials crossing partitions are unused. The held-out test partition is never used for training, model selection, threshold tuning, or feature selection.


\begin{table}[!t]
\centering
{\let\footnotesize\scriptsize
\caption{Dataset statistics for the speaker-disjoint development-derived partitions and the VoxCeleb1B sanity-check split.}
\label{tab:dataset_protocols}
}
\fontsize{8}{10}\selectfont
\setlength{\tabcolsep}{0pt}
\renewcommand{\arraystretch}{1.03}
\begin{tabular*}{\columnwidth}{@{\extracolsep{\fill}}llrrrrrr@{}}
\hline
Dataset & Split & Trials & Spk. & Tar. & Non & Same & Cross \\
\hline
\multirow{3}{*}{TidyVoiceX}
& Calib. & 5,372,011 & 485 & 2,541,166 & 2,830,845 & 2,568,780 & 2,803,231 \\
& Valid. & 1,169,874 & 162 & 810,257   & 359,617   & 619,607   & 550,267 \\
& Test   & 922,640   & 161 & 648,577   & 274,063   & 500,778   & 421,862 \\
\hline
\multirow{2}{*}{VoxCeleb1B}
& Calib. & 315,184 & 508 & 198,225 & 116,959 & 116,959 & 198,225 \\
& Test   & 292,066 & 267 & 206,062 & 86,004  & 86,004  & 206,062 \\
\hline
\end{tabular*}
\end{table}

VoxCeleb1B is used only as a no-language or score-only sanity check~\cite{nagrani2017voxceleb}. It is not used for the main metadata-driven embedding-fusion claim because language labels may be confounded with protocol construction~\cite{nam2023disentangled}.
The group-level protocol is reported only as secondary validation because its input structure differs from the trial-level AMECxSV setting.

\begin{table}[!t]
\centering
\caption{Comparison map for systems and controls.}
\label{tab:system_taxonomy}
\fontsize{8}{10}\selectfont
\setlength{\tabcolsep}{2pt}
\renewcommand{\arraystretch}{1.05}
\begin{tabular*}{\columnwidth}{@{\extracolsep{\fill}}>{\raggedright\arraybackslash}p{0.20\columnwidth}>{\raggedright\arraybackslash}p{0.34\columnwidth}>{\raggedright\arraybackslash}p{0.40\columnwidth}@{}}
\hline
Group & System & Input features \\
\hline
\textit{Basic-score} & Raw Single-Score & raw single score \\
 & Logistic calibration & single score \\
 & PAV / isotonic & single score \\
 & QMF calibration & single score + QMF duration feature \\
 & Language calibration & single score + $L_i$ \\
 & Reliability calibration & single score + $\mathbf{r}_i$ \\
\hline
\textit{Multi-score} & MultiScore-FC & score vector \\
 & MultiScore-ABS & score vector \\
\hline
\textit{Proposed} & AMEC-FC & score vector + $L_i$ + $\mathbf{r}_i$ \\
 & AMEC-ABS & score vector + $L_i$ + $\mathbf{r}_i$ \\
\hline
\textit{Control} & Strict Architecture-Matched Score-Only & score vector + constant metadata slots \\
 & Metadata Permutation Control & score vector + shuffled $L_i,\mathbf{r}_i$ \\
 & Metadata-Only Negative Control & $L_i$ + $\mathbf{r}_i$ only \\
\hline
\textit{Side check} & Group-Level Baseline & group evidence + uncertainty \\
 & Group-Level Abstention & group evidence + uncertainty \\
\hline
\end{tabular*}
\end{table}

\subsection{Compared Systems and Controls}
\label{subsec:compared_systems}
Table~\ref{tab:system_taxonomy} summarizes each compared system by group, name, and input features.
Throughout the results, MultiScore-FC denotes the score-vector full-coverage control, AMEC-FC denotes the proposed metadata-driven full-coverage AMECxSV variant, and the corresponding ``-ABS'' variants add the same validation-tuned posterior-confidence gate.

MultiScore-FC uses only the multi-front-end score vector, whereas AMEC-FC adds $L_i$ and $\mathbf{r}_i$ as calibration context. 
The metadata permutation, strict architecture-matched score-only, and metadata-only controls test whether the gain comes from aligned context rather than input size, MLP capacity, or metadata shortcuts. 
Group-level systems use a separate input structure and serve only as secondary checks.

\subsection{Implementation Details}
The six fixed internal embedding systems are SpeechBrain ECAPA-TDNN, WeSpeaker ResNet34, 3D-Speaker CAM++, 3D-Speaker ERes2NetV2, Microsoft WavLM Base, and Microsoft WavLM Base+~\cite{desplanques2020ecapa,ravanelli2021speechbrain,wang2023wespeaker,wang2023campp,chen2024eres2netv2,chen20243dspeaker,chen2021wavlm}. 
All embedding extractors and score backends are frozen.

For internal AMECxSV, six scores are concatenated with \texttt{same\_language}, \texttt{min\_duration}, \texttt{duration\_ratio}, and \texttt{short\_duration\_risk}. These 10 raw inputs expand to 112 features through score summaries, order statistics, pairwise score terms, score-metadata interactions, and metadata squares, then are standardized with training-set statistics. The expansion uses no labels, speaker identities, trial IDs, or trial-identity fields.

The AMECxSV head is a flat-vector MLP with two hidden layers of width 24. Each hidden layer uses Linear--LayerNorm--SiLU--Dropout with dropout 0.12~\cite{ba2016layer,srivastava2014dropout,ramachandran2017searching}, followed by a scalar output. It uses no separate score/metadata branches, residual connections, gating, or attention. Internal runs use AdamW~\cite{loshchilov2019decoupled} with learning rate $5.00\times10^{-4}$, weight decay $1/30$, batch size 131072, up to 300 epochs, validation $C_{\mathrm{llr}}$ scheduling with patience 10 and factor 0.50, early stopping patience 60, balanced binary cross-entropy weights, and validation-split affine calibration. The rich-feature MLP controls use fixed training seed 0 and are not a multi-seed robustness sweep.

External evaluation uses the official TidyVoice SimAM-ResNet34 system and the LI-MSV w2v-BERT 2.0 system as fixed front ends~\cite{farhadipour2026tidyvoicechallenge,qin2022simam,li2026language,barrault2023seamless,chung2021w2vbert}. 
We evaluate cached scores directly and train source-specific score-only and score+metadata MLP heads on the calibration split; these experiments test source-specific recalibration, not zero-shot reuse of one backend. The dual-score external head combines both score sources; external heads use hidden width 128, dropout 0.15, AdamW learning rate $5.00\times10^{-4}$, weight decay $1/10$, batch size 262144, and 120 epochs.
The best model is selected according to the lowest loss on the development set.

\subsection{Evaluation Metrics}
We report EER, $C_{\mathrm{llr}}$, and actDCF at $P_{\mathrm{tar}}\in\{0.001,0.01\}$ following standard ASV detection and calibration practice~\cite{martin1997det,brummer2006application,brummer2013bosaris}. EER measures ranking-based discrimination, $C_{\mathrm{llr}}$ measures calibrated log-likelihood-ratio quality, and actDCF evaluates calibrated decisions at fixed target priors. Appendix~\ref{sec:appendix_implementation_notes} adds implementation notes.

For abstention systems, accepted-trial EER, $C_{\mathrm{llr}}$, actDCF, and FAR are computed only over accepted trials and are reported together with coverage. These are therefore selective metrics rather than full-coverage metrics.

Numeric entries in the tables are reported with 95\% bootstrap confidence intervals obtained from 1,000 trial-level bootstrap resamples~\cite{efron1994bootstrap}, computed using an existing toolkit~\cite{Confidence_Intervals}. For the primary AMEC-FC versus MultiScore-FC comparison, we additionally report paired speaker-clustered bootstrap intervals over enrollment-speaker clusters.

\begin{table*}[!t]
\centering
\caption{Main TidyVoiceX-ASV trial-level results on the development-derived held-out partition. Rows with coverage below 1.00 report accepted-trial selective metrics, not full-coverage EER. Cells show estimates $\pm$ 95\% bootstrap CI half-widths.}
\label{tab:main_results}
\fontsize{8}{10}\selectfont
\setlength{\tabcolsep}{2pt}
\begin{tabular*}{\textwidth}{@{\extracolsep{\fill}}lccccc@{}}
\hline
System & Cov. & EER (\%) & $C_{\mathrm{llr}}$ & actDCF$_{0.001}$ & actDCF$_{0.01}$ \\
\hline
\multicolumn{6}{l}{\textit{Full-coverage decision systems}} \\
Raw Single-Score & $1.00\pm0.00$ & $3.60\pm0.04$ & $0.90\pm0.00$ & $1.00\pm0.00$ & $1.00\pm0.00$ \\
Global Logistic Calibration & $1.00\pm0.00$ & $3.60\pm0.04$ & $0.14\pm0.00$ & $0.73\pm0.00$ & $0.42\pm0.01$ \\
PAV / Isotonic Calibration & $1.00\pm0.00$ & $3.60\pm0.04$ & $0.14\pm0.00$ & $0.97\pm0.00$ & $0.38\pm0.01$ \\
Single-Score QMF Calibration & $1.00\pm0.00$ & $3.58\pm0.04$ & $0.14\pm0.00$ & $1.00\pm0.00$ & $0.54\pm0.01$ \\
Single-Score Language Calibration & $1.00\pm0.00$ & $3.52\pm0.04$ & $0.14\pm0.00$ & $0.77\pm0.01$ & $0.55\pm0.01$ \\
MultiScore-FC & $1.00\pm0.00$ & $2.15\pm0.03$ & $0.09\pm0.00$ & $0.64\pm0.05$ & $0.38\pm0.02$ \\
AMEC-FC & $1.00\pm0.00$ & $1.85\pm0.03$ & $0.07\pm0.00$ & $0.47\pm0.02$ & $0.28\pm0.01$ \\
\hline
\multicolumn{6}{l}{\textit{Selective decision systems, accepted-trial metrics}} \\
Single-Score Reliability Gate & $0.88\pm0.00$ & $3.86\pm0.04$ & $0.15\pm0.00$ & $0.96\pm0.00$ & $0.48\pm0.01$ \\
MultiScore-ABS & $0.81\pm0.00$ & $0.22\pm0.02$ & $0.02\pm0.00$ & $0.56\pm0.07$ & $0.22\pm0.02$ \\
AMEC-ABS & $0.80\pm0.00$ & $0.10\pm0.01$ & $0.01\pm0.00$ & $0.33\pm0.03$ & $0.10\pm0.01$ \\
\hline
\end{tabular*}
\end{table*}

\section{Results and Analysis}
This section reports controlled internal results, feature ablations, coverage-risk trade-offs, external fixed-front-end results, secondary checks, and limitations. Additional checks are in Appendices~\ref{sec:appendix_stat_metadata}--\ref{sec:appendix_implementation_notes}.

\subsection{Controlled Trial-Level Results}
Table~\ref{tab:main_results} summarizes the main TidyVoiceX-ASV trial-level results using descriptive system names.
Classical calibration mainly improves calibration and score scale rather than ranking. Logistic and isotonic calibration leave the Raw Single-Score EER essentially unchanged while substantially improving $C_{\mathrm{llr}}$ and operating-point costs.

The single-score metadata baselines are controls, not the main evidence for AMECxSV. Their gains are limited because metadata are applied to one score stream rather than to multi-score evidence.
AMEC-FC improves over MultiScore-FC across discrimination, calibration, and operating-point metrics on this split. This supports the use of aligned metadata as calibration context beyond ordinary multi-score fusion.

Trials sharing the same enrollment speaker are correlated. Therefore, for the primary full-coverage comparison between AMEC-FC and MultiScore-FC, we additionally performed 1,000 paired bootstrap replicates over 161 enrollment-speaker clusters.
AMEC-FC reduced EER by 0.30 percentage points relative to MultiScore-FC (95\% CI [$-0.51$, $-0.08$], paired $p=0.00$) and reduced $C_{\mathrm{llr}}$ by 0.01 (95\% CI [$-0.02$, $0.00$], paired $p=0.01$). 
actDCF also decreased significantly at both target priors; see Table~\ref{tab:supp_cluster_bootstrap}.
AMEC-ABS improves accepted-trial metrics relative to MultiScore-ABS at comparable coverage, but this should be interpreted as a selective-metric improvement rather than a full-coverage EER reduction.

\subsection{Metadata-Control Sanity Checks}

Table~\ref{tab:causal_controls} isolates the metadata effect. The matched score-only control keeps AMEC-FC's 10 raw input slots, 112-dimensional expansion, MLP architecture, parameter count, validation selection, and fixed training seed, but replaces metadata with constant placeholders. Its performance remains close to MultiScore-FC, shuffled metadata similarly removes the gain, and metadata-only is nearly unusable. These controls support interpreting aligned metadata as calibration context rather than added capacity or a metadata-label shortcut. Additional controls are in Appendix~\ref{sec:appendix_stat_metadata}.

\begin{table}[!b]
\centering
\caption{Causal-isolation controls for AMEC-FC. Values are estimates $\pm$ 95\% bootstrap CI half-widths.}\label{tab:causal_controls}
\fontsize{8}{10}\selectfont
\setlength{\tabcolsep}{2pt}
\begin{tabular*}{\columnwidth}{@{\extracolsep{\fill}}lcc@{}}
\hline
System & EER (\%) & $C_{\mathrm{llr}}$ \\
\hline
MultiScore-FC & $2.15\pm0.03$ & $0.09\pm0.00$ \\
AMEC-FC & $\mathbf{1.85\pm0.03}$ & $\mathbf{0.07\pm0.00}$ \\
Matched score-only & $2.14\pm0.03$ & $0.09\pm0.00$ \\
Shuffled metadata & $2.13\pm0.03$ & $0.09\pm0.00$ \\
Metadata-only & $46.06\pm0.11$ & $0.99\pm0.00$ \\
\hline
\end{tabular*}
\end{table}

Condition-stratified checks support the same bounded interpretation. AMEC-FC improves both same-language and cross-language subsets, and duration-ratio strata show EER reductions of 0.19, 0.42, and 0.22 percentage points for severe, moderate, and matched duration ratios, with $C_{\mathrm{llr}}$ reductions in all three groups. For minimum duration below 3 s, $C_{\mathrm{llr}}$ improves from 0.13 to 0.11, but EER changes from 2.62\% to 2.71\%, so the short-duration result is a calibration gain rather than a uniform ranking gain.

\subsection{Feature Ablation}
Table~\ref{tab:amec_ablation} reports the AMECxSV feature ablation under full-coverage and abstention settings.

\begin{table}[!b]
\centering
{\let\footnotesize\scriptsize
\caption{AMECxSV feature ablation with 95\% bootstrap CIs.}
\label{tab:amec_ablation}
}
\fontsize{8}{10}\selectfont
\setlength{\tabcolsep}{1pt}
\renewcommand{\arraystretch}{1.03}
\resizebox{\columnwidth}{!}{%
\begin{tabular}{@{}lcccc@{}}
\hline
\multicolumn{5}{c}{Full coverage} \\
\hline
Metric & \begin{tabular}{@{}c@{}}AMEC\\FC\end{tabular} & \begin{tabular}{@{}c@{}}Score-only\\FC\end{tabular} & \begin{tabular}{@{}c@{}}w/o\\Language\end{tabular} & \begin{tabular}{@{}c@{}}w/o\\Reliability\end{tabular} \\
\hline
EER (\%) & $1.85\pm0.03$ & $2.14\pm0.03$ & $2.06\pm0.03$ & $1.99\pm0.03$ \\
$C_{\mathrm{llr}}$ & $0.07\pm0.00$ & $0.09\pm0.00$ & $0.08\pm0.00$ & $0.08\pm0.00$ \\
actDCF$_{0.01}$ & $0.28\pm0.01$ & $0.37\pm0.02$ & $0.36\pm0.02$ & $0.31\pm0.01$ \\
\hline
\multicolumn{5}{c}{Abstention} \\
\hline
Metric & \begin{tabular}{@{}c@{}}AMEC\\ABS\end{tabular} & \begin{tabular}{@{}c@{}}Score-only\\ABS\end{tabular} & \begin{tabular}{@{}c@{}}w/o\\Language\end{tabular} & \begin{tabular}{@{}c@{}}w/o\\Reliability\end{tabular} \\
\hline
Cov. & $0.80\pm0.00$ & $0.80\pm0.00$ & $0.80\pm0.00$ & $0.81\pm0.00$ \\
Acc. EER (\%) & $0.10\pm0.01$ & $0.20\pm0.02$ & $0.18\pm0.02$ & $0.13\pm0.01$ \\
Acc. $C_{\mathrm{llr}}$ & $0.01\pm0.00$ & $0.02\pm0.00$ & $0.01\pm0.00$ & $0.01\pm0.00$ \\
FAR & \begin{tabular}{@{}c@{}}$(7.84\pm1.03)$\\$\times 10^{-4}$\end{tabular} & \begin{tabular}{@{}c@{}}$(17.30\pm1.55)$\\$\times 10^{-4}$\end{tabular} & \begin{tabular}{@{}c@{}}$(14.74\pm1.49)$\\$\times 10^{-4}$\end{tabular} & \begin{tabular}{@{}c@{}}$(10.33\pm1.14)$\\$\times 10^{-4}$\end{tabular} \\
\hline
\end{tabular}%
}
\end{table}

AMEC-FC and AMEC-ABS are best in their respective full-coverage and confidence-reject settings. Removing either language metadata or reliability metadata degrades performance. The no-language variant remains above the strict score-only control at full coverage (2.06\% versus 2.14\% EER), but it does not recover the full AMEC-FC result (1.85\% EER). Thus, duration metadata provide a language-blind fallback, whereas the full gain requires language-match metadata.

\subsection{Coverage--Risk Trade-off}
\begin{figure}[!t]
\centering
\includegraphics[width=\linewidth]{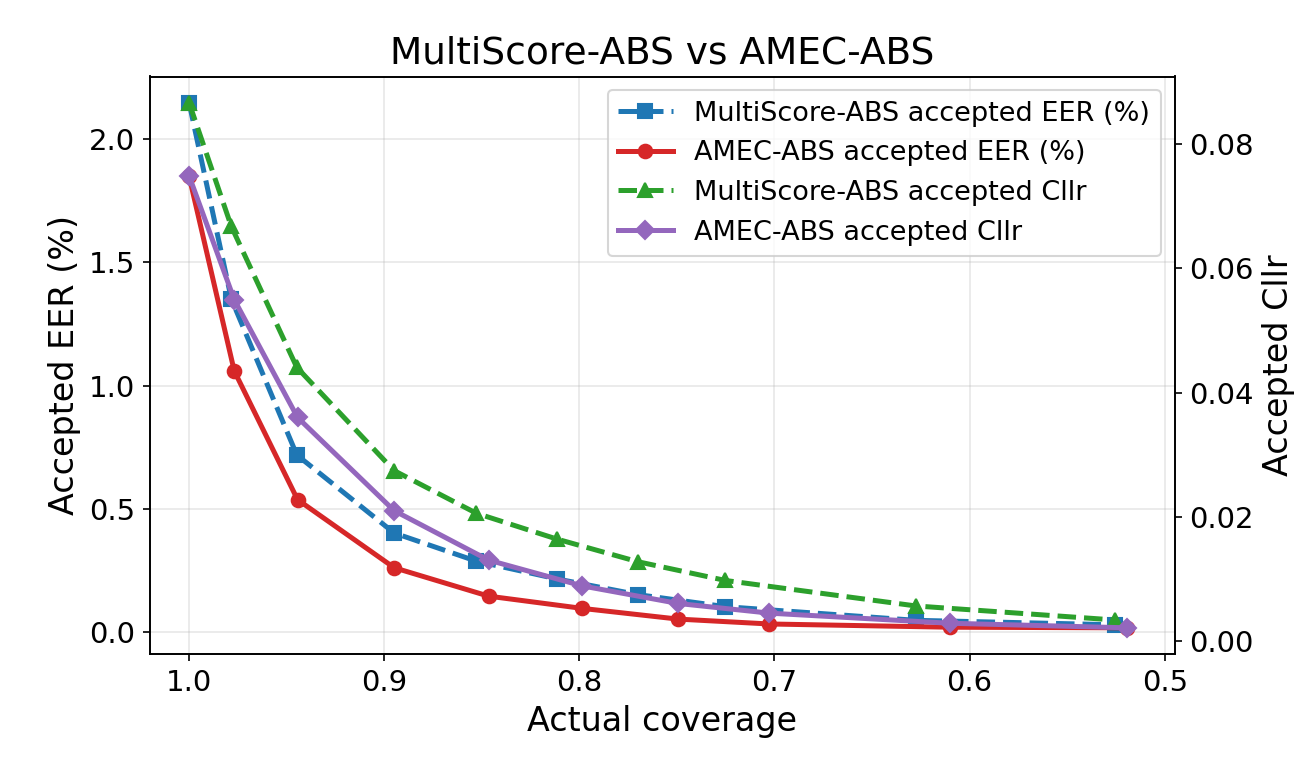}
\caption{Coverage-risk curve for AMEC-ABS. Lower coverage retains higher-confidence automatic decisions and reduces accepted-trial error.}
\label{fig:amec_coverage_curve}
\end{figure}

Figure~\ref{fig:amec_coverage_curve} shows AMEC-ABS coverage-risk behavior: at 0.80 coverage, accepted-trial EER is 0.10\%. Across the sweep, AMEC-ABS reduces EER AURC from 0.32 to 0.22, $C_{\mathrm{llr}}$ AURC from 0.02 to 0.01, and actDCF$_{0.01}$ AURC from 0.18 to 0.10 relative to MultiScore-ABS.

At the nominal 0.80 operating point, AMEC-ABS and MultiScore-ABS reject broadly similar target and non-target fractions: 20.42\%/19.54\% versus 19.25\%/17.87\%. 
Thus the accepted-trial gain is not driven by dropping only one class. Selective decisions require a fallback for rejected trials~\cite{geifman2017selective,ferrer2022failsafe}.

\subsection{External Fixed-Front-End Results}
Public leaderboard results are only informal scale references; all external experiments use the same development-derived speaker-disjoint held-out split. 
We evaluate the official TidyVoice SimAM-ResNet34 and LI-MSV w2v-BERT 2.0 score sources as fixed external front ends~\cite{farhadipour2026tidyvoicechallenge,qin2022simam,li2026language,barrault2023seamless,chung2021w2vbert}.

\begin{figure}[!t]
\centering
\includegraphics[width=\linewidth]{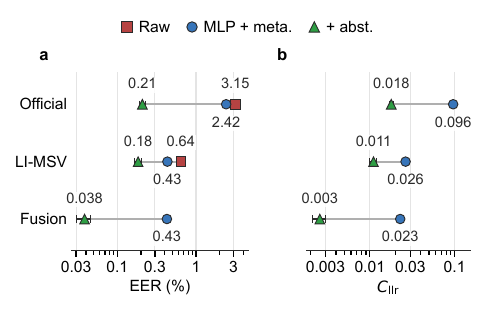}
\caption{External baseline comparison on the development-derived speaker-disjoint held-out split. Points show estimates, with 95\% bootstrap CIs where available, on log-scaled metric axes. The ``+ abst.'' points report accepted-trial metrics; the best result reaches 0.03\% accepted-trial EER at 0.79 coverage and is not a full-coverage or official-evaluation metric.}
\label{fig:external_tidyvoice_baselines}
\end{figure}

Figure~\ref{fig:external_tidyvoice_baselines} reports the external results. At full coverage, score+metadata heads reduce EER from 3.15\% to 2.42\% for Official, from 0.64\% to 0.43\% for LI-MSV, and from 0.63\% to 0.43\% for LI-MSV plus Official. The best external selective result is 0.03\% accepted-trial EER at 0.79 coverage for LI-MSV plus Official score+metadata. Additional external settings are in Appendix~\ref{sec:appendix_external_deployment}.

\subsection{Secondary Sanity Checks}
VoxCeleb1B and group-level experiments are secondary checks, not main evidence. 
VoxCeleb1B is used only as a no-language or score-only sanity check because language metadata may be confounded with protocol construction~\cite{nagrani2017voxceleb,nam2023disentangled}; AMECxSV-like calibration changes EER from 2.02\% to 1.90\%, with comparable abstention variants. 
The group-level protocol uses a separate multi-enrollment input structure, so its abstention gains support the confidence mechanism only as secondary evidence.

\subsection{Limitations and Negative Findings}
Several score transforms and fusion architectures were evaluated, including AS-Norm-, WCCN-, LDA-, and Mahalanobis-style variants~\cite{matejka2017analysis,hatch2006wccn,dehak2011frontend,mahalanobis1936generalized}, but did not consistently improve over direct cosine-score fusion or the flat-vector MLP. AMECxSV remains a score-derived backend: it does not solve embedding-level language mismatch, depends on upstream score quality, and needs a fallback policy for abstained trials.

The full score+metadata setting requires duration metadata and either oracle or predicted language metadata at inference time. In language-blind protocols, the language-match feature must be removed or supplied by a validated LID module; see Appendix~\ref{sec:appendix_external_deployment}.

Benefits are condition- and source-dependent: short-duration trials improve in $C_{\mathrm{llr}}$ but not EER, and external gains vary by score source. Validation is reused for early stopping, scheduling, affine calibration, and thresholds, but the held-out test partition is untouched. Because the split is development-derived and excludes cross-partition non-target trials, results are controlled development evidence, not official challenge scores.

\section{Conclusion and Future Work}
This paper presented AMECxSV, a metadata-driven embedding-fusion calibration backend for fixed X-lingual ASV front ends under metadata-available trial conditions.
On a speaker-disjoint TidyVoiceX-ASV development split, AMEC-FC improved the internal MultiScore-FC full-coverage control, and external comparisons showed source-dependent score+metadata gains. 
Confidence rejection reduced accepted-trial risk at stated coverage. 
Claims are limited to calibration and confidence control for tested fixed score sources with available or ablated metadata, not representation learning, language-blind or official challenge scoring, or untested transfer.

\bibliographystyle{IEEEtran}
\bibliography{references}

\clearpage
\onecolumn

\appendices

\section{Statistical and Metadata-Control Evidence}
\label{sec:appendix_stat_metadata}
These retained controls support the main metadata-calibration interpretation without adding long implementation tables.

\begin{table}[H]
\centering
\caption{Speaker-clustered paired bootstrap for AMEC-FC versus MultiScore-FC on the TidyVoiceX-ASV development-derived held-out test partition. Differences are AMEC-FC minus MultiScore-FC, so negative values are improvements for lower-is-better metrics.}
\label{tab:supp_cluster_bootstrap}
\setlength{\tabcolsep}{4pt}
\resizebox{\columnwidth}{!}{%
\begin{tabular*}{\textwidth}{@{\extracolsep{\fill}}lccccc@{}}
\hline
Metric & MultiScore-FC & AMEC-FC & Difference & 95\% CI & Paired $p$ \\
\hline
EER (\%) & 2.15 & 1.85 & -0.30 & [-0.51, -0.08] & 0.00 \\
$C_{\mathrm{llr}}$ & 0.09 & 0.07 & -0.01 & [-0.02, 0.00] & 0.01 \\
actDCF$_{0.001}$ & 0.64 & 0.47 & -0.18 & [-0.37, -0.02] & 0.02 \\
actDCF$_{0.01}$ & 0.38 & 0.28 & -0.09 & [-0.20, -0.02] & 0.01 \\
\hline
\end{tabular*}
}
\end{table}

\begin{table}[H]
\centering
\caption{Same-feature linear and MLP controls on the held-out split. Linear rows use the same calibration split, validation $C_{\mathrm{llr}}$ selection, standardization protocol, and validation-split affine calibration as the MLP rows. Metric cells show estimates $\pm$ 95\% bootstrap CI half-widths.}
\label{tab:reviewer_linear_controls}
\setlength{\tabcolsep}{4pt}
\resizebox{\columnwidth}{!}{%
\begin{tabular*}{\textwidth}{@{\extracolsep{\fill}}llcccc@{}}
\hline
System & Feature set & Head & EER (\%) & $C_{\mathrm{llr}}$ & actDCF$_{0.01}$ \\
\hline
MultiScore-FC & score-only summaries & MLP & $2.15\pm0.03$ & $0.09\pm0.00$ & $0.38\pm0.02$ \\
AMEC-FC & scores + language/duration metadata & MLP & $1.85\pm0.03$ & $0.07\pm0.00$ & $0.28\pm0.01$ \\
Linear score-only raw & six raw scores & linear & $2.44\pm0.03$ & $0.10\pm0.00$ & $0.39\pm0.02$ \\
Linear score+metadata raw & six scores + $L_i$ + duration metadata & linear & $2.08\pm0.03$ & $0.08\pm0.00$ & $0.29\pm0.01$ \\
Linear matched score-only & expanded scores + constant metadata slots & linear & $2.31\pm0.03$ & $0.09\pm0.00$ & $0.36\pm0.02$ \\
Linear AMEC-expanded & expanded score-metadata interactions & linear & $1.98\pm0.03$ & $0.08\pm0.00$ & $0.28\pm0.01$ \\
\hline
\end{tabular*}
}
\end{table}

\section{External and Deployment-Oriented Checks}
\label{sec:appendix_external_deployment}
External rows use the same development-derived held-out split as the main text. Rows with coverage below 1.00 are accepted-trial selective metrics.

\begin{table}[H]
\centering
\caption{Merged external fixed-front-end results, combining raw-score, score-only, score+metadata, and confidence-gated settings. Metric cells show estimates $\pm$ 95\% bootstrap CI half-widths where defined; rows with coverage below 1.00 are accepted-trial selective metrics.}
\label{tab:supp_external_score_only}
\setlength{\tabcolsep}{3pt}
\resizebox{\columnwidth}{!}{%
\begin{tabular*}{\textwidth}{@{\extracolsep{\fill}}lllcccc@{}}
\hline
Source & Setting & Scope & Coverage & EER (\%) & $C_{\mathrm{llr}}$ & actDCF$_{0.01}$ \\
\hline
Official & Raw score & full & $1.00\pm0.00$ & $3.15\pm0.04$ & n/a & n/a \\
Official & Score-only MLP & full & $1.00\pm0.00$ & $3.15\pm0.04$ & $0.12\pm0.00$ & $0.56\pm0.01$ \\
Official & Score+metadata MLP & full & $1.00\pm0.00$ & $2.42\pm0.03$ & $0.10\pm0.00$ & $0.54\pm0.01$ \\
Official & Score+metadata gate & accepted & $0.79\pm0.00$ & $0.49\pm0.02$ & $0.03\pm0.00$ & $0.37\pm0.01$ \\
\hline
LI-MSV & Raw score & full & $1.00\pm0.00$ & $0.64\pm0.02$ & n/a & n/a \\
LI-MSV & Score-only MLP & full & $1.00\pm0.00$ & $0.64\pm0.02$ & $0.04\pm0.00$ & $0.44\pm0.02$ \\
LI-MSV & Score+metadata MLP & full & $1.00\pm0.00$ & $0.43\pm0.02$ & $0.03\pm0.00$ & $0.40\pm0.02$ \\
LI-MSV & Score+metadata gate & accepted & $0.78\pm0.00$ & $0.17\pm0.02$ & $0.01\pm0.00$ & $0.17\pm0.01$ \\
\hline
Fusion & Score-only MLP & full & $1.00\pm0.00$ & $0.63\pm0.02$ & $0.03\pm0.00$ & $0.41\pm0.02$ \\
Fusion & Score+metadata MLP & full & $1.00\pm0.00$ & $0.43\pm0.02$ & $0.02\pm0.00$ & $0.26\pm0.01$ \\
Fusion & Score+metadata gate & accepted & $0.79\pm0.00$ & $0.03\pm0.01$ & $0.00\pm0.00$ & $0.03\pm0.01$ \\
\hline
\end{tabular*}
}
\end{table}

\begin{table}[H]
\centering
\caption{Predicted language metadata experiment. The predicted-language row replaces oracle $L_i$ with $\hat{L}_i$ constructed from a frozen LID model. Score inputs, duration metadata, training schedule, validation affine calibration, and held-out test protocol are unchanged. Metric cells show estimates $\pm$ 95\% bootstrap CI half-widths when available.}
\label{tab:appendix_predicted_lid}
\setlength{\tabcolsep}{4pt}
\resizebox{\columnwidth}{!}{%
\begin{tabular}{llccccc}
\hline
System & Language metadata source & Test $L$ agreement & EER (\%) & $C_{\mathrm{llr}}$ & actDCF$_{0.001}$ & actDCF$_{0.01}$ \\
\hline
MultiScore-FC & none & n/a & $2.15\pm0.03$ & $0.09\pm0.00$ & $0.64\pm0.05$ & $0.38\pm0.02$ \\
AMEC-FC & oracle $L_i$ & 1.00 & $1.85\pm0.03$ & $0.07\pm0.00$ & $0.47\pm0.02$ & $0.28\pm0.01$ \\
AMEC-FC & frozen-LID predicted $\hat{L}_i$ & 0.78 & $2.07\pm0.03$ & $0.08\pm0.00$ & $0.74\pm0.02$ & $0.36\pm0.02$ \\
AMEC-FC w/o language & duration only & n/a & $2.06\pm0.03$ & $0.08\pm0.00$ & n/a & $0.36\pm0.02$ \\
\hline
\end{tabular}
}
\end{table}

SpeechBrain's VoxLingua107 ECAPA LID model was run once per unique utterance and was not trained or fine-tuned on ASV labels, speaker identities, or trial identities. The frozen-LID row bounds the deployment claim: AMECxSV can consume predicted language metadata, but reliable LID is needed at stringent operating points.

\section{Implementation Notes}
\label{sec:appendix_implementation_notes}
The MLP output $u_i$ is affine-calibrated as $\ell_i=\alpha u_i+\beta$, with $\alpha,\beta$ fitted on validation scores by class-balanced binary cross-entropy; $C_{\mathrm{llr}}$ is then computed on $\ell_i$, treated directly as the log-likelihood-ratio score. Reported actDCF uses fixed application thresholds $\log((1-P_{\mathrm{tar}})/P_{\mathrm{tar}})$ and is not minDCF unless minDCF is explicitly named. Abstention metrics are computed only on accepted trials and must be read with coverage. Raw-score $C_{\mathrm{llr}}$ is omitted because uncalibrated raw scores are not calibrated log-likelihood-ratio scores and no calibrated logit is trained for those rows.

\end{document}